\documentclass[prb,twocolumn,showpacs,superscriptaddress]{revtex4}
\pdfoutput=1 
\usepackage{epsfig}
\usepackage{amsmath,amssymb}
\usepackage{mathtools}
\usepackage{graphicx}
\usepackage{footnote}
\usepackage[caption=false]{subfig}
\usepackage{tabularx}
\usepackage[normalem]{ulem}
\usepackage{color}
\definecolor{darkblue}{rgb}{0,0,0.35}
\usepackage{array}
\usepackage{placeins}
\usepackage[all]{xy}

\newcommand{\ceq}[1]{Eq.~(\ref{#1})}
\newcommand{\cfg}[1]{Fig.~\ref{#1}}
\newcommand{\overbar}[1]{\mkern 1.5mu\overline{\mkern-1.5mu#1\mkern-1.5mu}\mkern 1.5mu}

\newcommand*\sumint{\,\,\, \mathclap{\displaystyle\int}\mathclap{\textstyle\sum} \;\;\;}

\usepackage{tikz}
\usetikzlibrary{arrows,shapes,positioning}
\usetikzlibrary{decorations.markings}
\usetikzlibrary{calc}
\usetikzlibrary{external}
\tikzstyle arrowstyle=[scale=1]

\newcommand*{\halfwaydir}{0.5*\pgfdecoratedpathlength+4pt}
\newcommand*{\halfwayrev}{0.5*\pgfdecoratedpathlength-4pt}
\tikzstyle directed=[postaction={decorate,decoration={markings,
    mark=at position \halfwaydir with {\arrow[arrowstyle]{stealth}}}}]
\tikzstyle reverse directed=[postaction={decorate,decoration={markings,
    mark=at position \halfwayrev with {\arrowreversed[arrowstyle]{stealth}}}}]

\newcommand{\uu}{{\uparrow\uparrow}}

\newcommand{\ud}{{\uparrow\downarrow}}
\newcommand{\du}{{\downarrow\uparrow}}
\newcommand{\udb}{{\overbar{\uparrow\downarrow}}}

\newcommand{\udud}{{\uparrow\downarrow\uparrow\downarrow}}
\newcommand{\dudu}{{\downarrow\uparrow\downarrow\uparrow}}
\newcommand{\uddu}{{\uparrow\downarrow\downarrow\uparrow}}
\newcommand{\duud}{{\downarrow\uparrow\uparrow\downarrow}}
\newcommand{\uuuu}{{\uparrow\uparrow\uparrow\uparrow}}
\newcommand{\uudd}{{\uparrow\uparrow\downarrow\downarrow}}

\begin{document}

\title{Continuous-time quantum Monte Carlo calculation of multi-orbital vertex asymptotics}

\author{Josef~Kaufmann}
\affiliation{\small\em Institute for Solid State Physics, TU Wien, 1040 Vienna, Austria}
\author{Patrik~Gunacker}
\affiliation{\small\em Institute for Solid State Physics, TU Wien, 1040 Vienna, Austria}
\author{Karsten~Held}
\affiliation{\small\em Institute for Solid State Physics, TU Wien, 1040 Vienna, Austria}

\date{\small\today}
\begin{abstract}
We derive the equations for calculating the high-frequency asymptotics of the local two-particle vertex function for a multi-orbital impurity model. These relate the  asymptotics for a general local interaction to   equal-time two-particle Green's functions, which we sample
using  continuous-time quantum Monte Carlo simulations with a worm algorithm.
As specific examples we study the  single-orbital Hubbard model and  the three $t_{2g}$ orbitals of SrVO$_3$  within dynamical mean field theory (DMFT). We
demonstrate how the knowledge of the  high-frequency asymptotics reduces the statistical uncertainties of the vertex and further eliminates finite box size
effects. The proposed method benefits the calculation of non-local susceptibilities in  DMFT and diagrammatic extensions of DMFT.
\end{abstract}
\maketitle

\section{Introduction} \label{sec:Intro}
Strong electronic correlations are driving various properties of heavy fermion compounds, including Mott metal-to-insulator transitions,~\cite{Mott,Georges} magnetic phase transitions~\cite{Lohneysen,RohringerCritical}, and quantum critical points.~\cite{Schroder2000,SchaeferQCP}
While Mott transitions  can be described in terms of one-particle spectral functions only, the physics of the latter two is related to two-particle susceptibilities. 
Indeed, charge and magnetic susceptibilities are of primary interest when theoretical results 
are compared to experiments, but their computation in interacting systems is in general very costly.~\cite{LeBlanc}

Typically, the Hubbard model~\cite{Hubbard} is employed when investigating strong electronic correlations from the theoretical side. 
This model has been solved successfully within the dynamical mean field theory (DMFT)~\cite{Metzner,Georges:1992,Kotliar_dmft,Held} which corresponds 
to a  purely local self-energy. For determining this local self-energy,
DMFT maps the Hubbard model onto an auxiliary single-impurity Anderson model (AIM),~\cite{Anderson} which can be solved numerically.
Nowadays, a vast amount of impurity solvers exist, each having its particular strengths and weaknesses.\cite{Hirsch,Sakai,Bulla,Karski,Ganahl,Caffarel,Zgid}
A noteworthy group of impurity solvers includes the continuous-time quantum Monte Carlo (CT-QMC) methods, which can treat impurities with many degrees of freedom, general interactions, and continuous bath dispersions.~\cite{Rubstov_ct_int,Werner_qmc,Werner,Gull_aux,Gull} 
These algorithms are capable of calculating finite-temperature correlation functions (i.e. one- and two-particle Green's functions), which directly relate to the aforementioned susceptibilities and to vertex functions, respectively.

While DMFT is  exact in infinite spatial dimensions, the theory is often used as an approximation for finite dimensional systems.
In this case,  correlations that are non-local in space may emerge. There are several approaches which contain the local DMFT correlations but extend it for also including non-local ones.
The extensions of DMFT are grouped into cluster methods, which enlarge the AIM to multiple impurities or, alternatively, methods which diagrammatically improve upon DMFT. 
Promising diagrammatic extensions in this context include the dynamic vertex approximation ($\mathrm{D}\Gamma\mathrm{A}$),~\cite{Toschi} the dual fermion method (DF),~\cite{Rubtsov} the one-particle irreducible approach (1PI),~\cite{Rohringer_1PI} 
the DMFT to functional renormalization group ($\mathrm{DMF}^2\mathrm{RG}$)\cite{Taranto} and the quadruply irreducible local expansion (QUADRILEX).\cite{AryalQuad}
Although these methods follow in general quite different philosophies, they all rely on the knowledge of the local two-particle susceptibility or vertex function. These vertex functions have two incoming and two outgoing lines so that they depend on three frequencies [exploiting energy conservation] and two spin combinations [exploiting SU(2) symmetry].\cite{Rohringer} For multi-orbital calculations there are, on top of this, various combinations of the orbital degrees of freedom.

Albeit in principle straight-forward, it is  a very challenging task to extract the local multi-orbital two-particle-susceptibility of the AIM within large frequency-boxes. 
This can be traced back to the high computational resources in computing, storing and processing the two-particle object. 
Only recently the local two-particle correlation function with its complete frequency structure was obtained for SrVO$_3$ with SU(2)-symmetric interaction.~\cite{GallerDGA}
In order to overcome this limitation, contemporary attempts include approximating the asymptotic frequency behavior. 
The two main pieces of work in this direction are: (i) extracting the high-frequency asymptotics 
of the local two-particle  vertex function $\Gamma_{ph}$ that is irreducible in the particle-hole channel 
by approximating it by a certain sub-class of {\em single-frequency} susceptibility functions,~\cite{Kunes} 
(ii) extracting the complete high-frequency asymptotics of the full  vertex $F$ 
through all asymptotically contributing diagrams within the so-called kernel approximations, 
which include {\em one and two-frequency} kernel functions.~\cite{LiParquet,Wentzell}
While (ii) is not limited to a specific sub-class of diagrams (i.e. particle-hole, particle-particle ...) 
and yields the full asymptotics of the vertex, the derivation currently only exists for the single-orbital case. 
Approach (i), on the other hand, was implemented for multi-orbital systems in Ref.~\onlinecite{Kunes}, 
and successfully applied for calculating the $\omega=0$ susceptibility in DMFT,
but not for generalized susceptibilities or diagrammatic extensions of DMFT.   
For calculating $\Gamma_{ph}$, Ref.~\onlinecite{Kunes} also introduced an efficient implementation 
of the inversion of the Bethe-Salpeter equation. This has been extended to arbitrary 
channels and $\omega\neq 0$ in Ref.~\onlinecite{Tagliavini}.

In this paper we will follow approach (ii) in order to avoid divergences in the local two-particle irreducible 
vertex function~\cite{Schaefer,SchaeferNonPert} and further to include all local physics by considering all relevant diagrams. 
Since the kernel approximations are originally formulated for the vertex function 
instead of the susceptibility or correlation function, in this work we outline how to extract the kernel approximations from the correlation functions. 
Prior to this work, the kernel functions were approximated from the local two-particle vertex function itself by scanning the asymptotic region and 
employing this information for the functional renormalization group (fRG) flow~\cite{Wentzell}
and for the self-consistent solution of the parquet equations~\cite{LiParquet}. 
This approach is not suitable for quantum Monte Carlo algorithms due to the intrinsic statistical uncertainty.
Here, we demonstrate a method which directly allows us to measure the correlation functions related 
to the kernel functions with impurity solvers, such as CT-QMC or, 
in principle, any other type of impurity solver that is based on a  Green's function formalism. 
We further extend the kernel approximations by deriving the expressions for multi-orbital systems with general local interactions. 

Let us emphasize that the hybridization expansion (CT-HYB)~\cite{Werner} is the method of choice when dealing with the multi-orbital AIM at finite temperature and non-density-density interaction. 
We use a worm algorithm recently introduced to CT-HYB,~\cite{Gunacker,Gunacker16} to measure 
one- and two-time two-particle correlation functions, which are then transformed into the kernel functions.
Combining the sampling power of CT-HYB with the improvements due to the asymptotical structure 
allows us to access local physics of multi-orbital systems and especially of materials
with strongly reduced statistical uncertainty. \footnote{We point out that it is very common to approximate the high-frequency asymptotics of one-particle quantities such as the self-energy.
This is usually achieved by a frequency expansion in terms of moments. The zeroth and first moment directly follow from the one- and two-particle density matrices.\cite{Wang}}

In Section~\ref{sec:Theory} we present the theoretical foundation required for a rigorous definition of the multi-orbital kernel approximations.
Starting from the two-particle Green's function, we define the correlation functions, the susceptibilities and the vertex functions.
We further define the concepts of reducibility and irreducibility of two-particle quantities, respectively.
We show the local formulation of the parquet equations and the  necessary frequency representations.
In order to establish the connection between correlation functions and kernel approximations, we define  in Section~\ref{sec:Derivation} the equal-time susceptibilities and the corresponding multi-orbital kernel approximations. We further define the parameterization of the asymptotical structure and its connection to
the full vertex function.
We briefly present what modifications of the worm algorithm are necessary in Section~\ref{sec:Worm}, analyze the numerical effort,
and present a summary of the steps needed to calculate the Kernel functions. 
In Section~\ref{sec:Results} we apply the method to the single-orbital Hubbard model and benchmark our approach against results obtained from exact diagonalization (ED).
In a second step, we show results for the multi-orbital case, by calculating the asymptotical structure of SrVO$_3$, and outline the improvement with respect to the direct measurement of the two-particle correlation function. 
In Section~\ref{sec:Conclusion} we summarize our method in terms of its strengths and its prospective applications.
Our frequency conventions and additional derivations for the atomic limit are given in the Appendix.

\section{Hamiltonian and theoretical background}\label{sec:Theory}
In this paper, we consider the multi-orbital  AIM (which in DMFT is calculated self-consistently \cite{Georges,Held}):
\begin{multline}
\label{eq:anderson}
H = \frac{1}{4} \sum_{ijkl} U_{ijkl}^{\vphantom{dagger}} d^\dagger_{i} d^\dagger_{j} d^{\vphantom{dagger}}_{l} d^{\vphantom{dagger}}_{k} + \sum_{i} \tilde{\varepsilon}^{\vphantom{dagger}}_{i} d_{i}^\dagger d_{i} ^{\vphantom{dagger}}+ \\
+ \sum_{K i} \varepsilon^{\vphantom{dagger}}_{K i} c^\dagger_{Ki} c^{\vphantom{dagger}}_{Ki}
+ \sum_{K ij} \left[ V_{K}^{ij} c^\dagger_{K i} d^{\vphantom{dagger}}_{j} + ( V_{K}^{j i} )^* d^\dagger_i c^{\vphantom{dagger}}_{K j}\right] \;
\end{multline}
Here, $d_{i}$ ($d_{i}^\dagger$) is the annihilation (creation) operator of a fermion with spin-orbital flavor $i$,
$c_{Ki}$ ($c_{Ki}^{\dagger}$) is the annihilation (creation) operator of an electron with impurity flavor $i$
in the non-interacting bath and $K$ sums over the remaining bath degrees of freedom (e.~g.~the momentum $\mathbf{k}$).
The local impurity is described by a local one-particle potential $\tilde{\varepsilon}_i$ (e.g. the crystal field),
the fully anti-symmetrized interaction matrix $U_{ijkl}$, the bath dispersion $\varepsilon_{Ki}$, and the hybridization strength $V_{K}^{ij}$.

The $n$-particle Green's function of a local impurity in imaginary-time reads:
\begin{multline}
\label{eq:ngreen}
 G_{i_1 i_2 \ldots i_{2n-1} i_{2n}} (\tau_1, \tau_2, \ldots,\tau_{2n-1},\tau_{2n}) = \\ 
 (-1)^n \langle T_\tau d^{\vphantom{dagger}}_{i_1} (\tau_1) d_{i_2}^\dagger (\tau_2) \ldots d^{\vphantom{dagger}}_{i_{2n-1}} (\tau_{2n-1}) d_{i_{2n}}^\dagger (\tau_{2n}) \rangle,
\end{multline}
where $d_{i}(\tau_i)$ ($d_{i}^\dagger(\tau_i)$) are now the imaginary-time dependent annihilation (creation) operators at (imaginary) time $\tau_i$.
Further, $T_\tau$ is the imaginary-time ordering operator, and $\langle \ldots \rangle = (\text{Tr} e^{\beta H} \ldots)/Z $ the thermal expectation value at temperature $T$ ($\beta=1/T$), $Z$ is the partition function.
Expanding ~\ceq{eq:ngreen} into a perturbation series and decomposing it 
according to Wick's theorem yields all possible connected and disconnected Feynman diagrams. 
Distinguishing between disconnected and connected diagrams allows us to classify 
the $n$-particle Green's function into the $2n$-point correlation function 
and the subset of connected diagrams into $n$-particle vertex function.

At the two-particle level the Green's function decomposes into two disconnected parts, 
usually referred to as straight and cross terms, and a fully connected part:
\begin{multline}
\label{eq:2green} 
 G_{ijkl}(\tau_1, \tau_2, \tau_3, \tau_4) = G_{ij}(\tau_1, \tau_2) G_{kl}(\tau_3, \tau_4) \\
 \underbrace{-G_{il}(\tau_1, \tau_4) G_{kj}(\tau_2, \tau_3) + \chi^{\text{c}}_{ijkl}(\tau_1,\tau_2,\tau_3,\tau_4)}_{\equiv \chi_{ijkl}}.
\end{multline}
The cross term and the connected diagrams are 
further grouped into the generalized 4-point susceptibility $\chi_{ijkl}$.
\begin{figure}
\includegraphics{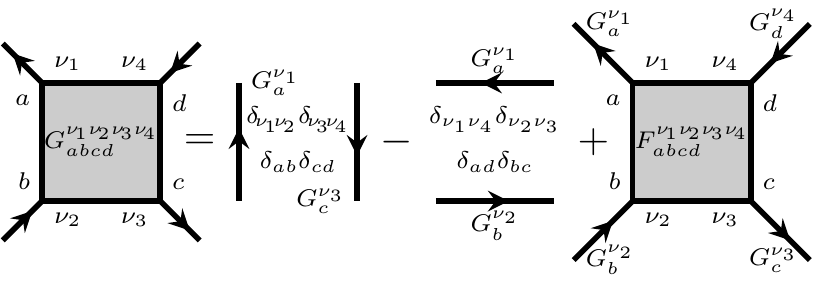}
\caption{\label{fig:g2chi} Decomposition of the two-particle Green's function into disconnected parts
  and a connected part. 
  }
\end{figure}
The two-particle vertex function $F_{mnop}$ now follows from the subset of connected diagrams by amputating the outer legs (one-particle Green's functions):
\begin{multline}
\label{eq:2F} 
\chi^{\text{c}}_{ijkl}(\tau_1,\tau_2,\tau_3,\tau_4) = - \sumint G_{im}(\tau_1, \tau_5) G_{nj}(\tau_6, \tau_2) \times \\
 F_{mnop}(\tau_5, \tau_6, \tau_7, \tau_8) G_{ko}(\tau_3, \tau_7) G_{pl}(\tau_8, \tau_4),
\end{multline}
where we integrate/sum over all internal imaginary time/spin-orbital degrees of freedom. 
That is, the $n$-particle vertex functions are defined without outer legs, 
whereas $n$-particle Green's functions and $2n$-point susceptibilities are defined with outer legs attached.

For any two-particle object considered in the following it is often useful to consider the Matsubara frequency representation, instead of the imaginary-time representation:
\begin{multline}
 \mathcal{A}^{\nu_1  \nu_2  \nu_3  \nu_4}_{ijkl} = \int_0^\beta  \mathrm{d}\tau_1 \mathrm{d} \tau_2 \mathrm{d} \tau_3 \mathrm{d} \tau_4 \times \\
 e^{i(\nu_1 \tau_1 - \nu_2 \tau_2 + \nu_3 \tau_3 -\nu_4 \tau_4)} 
 \mathcal{A}_{ijkl}(\tau_1,\tau_2,\tau_3,\tau_4),
\end{multline}
where $\mathcal{A}\in\{G,\chi,F\}$ and $\nu_i=(2n+1)\pi/\beta$ are the discrete fermionic Matsubara frequencies. The decomposition of the correlation function into disconnected parts and a fully connected part in Matsubara frequencies
is illustrated in \cfg{fig:g2chi}.
The back-transform is defined as:
\begin{multline}
\label{eq:back_transform}
 \mathcal{A}_{ijkl}(\tau_1,\tau_2,\tau_3,\tau_4) = \frac{1}{\beta^4} \times\\
 \sum_{\nu_1,\nu_2,\nu_3,\nu_4} e^{-i(\nu_1 \tau_1 - \nu_2 \tau_2 + \nu_3 \tau_3 -\nu_4 \tau_4)} \mathcal{A}^{\nu_1  \nu_2  \nu_3  \nu_4}_{ijkl}.
\end{multline}
When setting a single time-argument to zero only the frequency summation (without the exponential function) remains. 
This already implies that contracting two legs by a Matsubara frequency sum relates to setting the respective time differences to zero (which usually appear in mixed bosonic-fermionic frequency representations) thus resulting in an
equal-time object.

The time-translational symmetry inherent to the $n$-particle Green's function in imaginary time 
converts to an energy conservation in Matsubara frequency space. 
\begin{equation}
\label{eq:energy-conservation}
\nu_1 + \nu_3 = \nu_2 + \nu_4
\end{equation}
Consequently, it is sometimes more useful to assume a mixed bosonic-fermionic frequency representation with two fermionic and one bosonic frequency.
Each reducible channel introduced in the next section has its own natural frequency representation. The mapping between the four-frequency notation and the three-frequency notation that is natural in each channel is given in the Appendix \ref{sec:freq-not}.

When considering vertex functions in terms of Feynman diagrams it is useful to define the concept of reducibility. 
Here,  $n=1,2$-particle irreducible means that the vertex  cannot be separated into two or more parts by cutting $n$ Green's function lines.
At the one-particle level, the one-particle irreducible vertex can be obtained from the Dyson equation and is usually referred to as self-energy $\Sigma$.
At the two-particle level, it is necessary to consider reducibility more carefully. 
The two-particle vertex function $F$ is one-particle irreducible, however, it is not two-particle irreducible.

The (local) Parquet equations \cite{DeDominicis1962,DeDominicis1964,Bickers2004}decompose the two-particle vertex function $F$  into irreducible and reducible components:
\begin{equation}
\label{eq:parquet} 
F_{ijkl} = \Lambda_{ijkl} + \Phi^{ph}_{ijkl} + \Phi^{\overbar{ph}}_{ijkl} + \Phi^{pp}_{ijkl},
\end{equation}
where $\Lambda$ is the fully two-particle irreducible vertex function, and $\Phi^{ph},\Phi^{\overbar{ph}},\Phi^{pp}$ are the two-particle reducible vertex in the particle-hole ($ph$), the particle-hole transversal ($\overbar{ph}$) and the particle-particle ($pp$) channel.
In~\ceq{eq:parquet} we have omitted the time/frequency dependence of each quantity.
The subset of two-particle irreducible diagrams in a given channel $\ell=\{ph,\overbar{ph},pp\}$ is 
acquired by subtracting the reducible diagrams from the full vertex $F$, i.~e.
\begin{equation}
\Gamma^{\ell}_{ijkl} = F_{ijkl} - \Phi^{\ell}_{ijkl}.
\end{equation}
Constructing the reducible vertex functions as ladders leads to the Bethe-Salpeter equation
\begin{equation}
\label{eq:bsalpeter} 
\Gamma^{\ell} = F - \int \Gamma^\ell (GG)^\ell F.
\end{equation}
The asymptotic form of the two-particle irreducible vertex in the $ph$-channel
is calculated elsewhere,~\cite{Kunes} we focus on the full vertex $F$.

\section{Asymptotical structure of the local vertex}\label{sec:Derivation}
\subsection{Motivation}
In the following we derive the high-frequency asymptotics of the full                    two-particle vertex function $F$.
Alternatively, and in a very similar manner, one may derive the asymptotical behavior of the two-particle Green's function 
or the generalized susceptibility. The former, however, is superior, because contrary to the susceptibility,
the vertex can be parameterized very efficiently in its high-frequency region.\\
In order to describe this high-frequency asymptotics, we reiterate that 
outside of the low-frequency region only one contribution, the constant
background, originates from the two-particle-irreducible vertex $\Lambda$.\cite{Wentzell,Rohringer}
The remaining high-frequency-structures are contained in the vertices $\Phi^{\ell}$
reducible in channel $\ell$
and can be parameterized through much simpler one- and two-frequency objects,
coined Kernel-1 and Kernel-2 functions.\cite{Wentzell,Rohringer}
The constant background can be identified as the bare vertex
$U_{abcd}$ shown in \cfg{fig:umatrix}, which is the lowest-order term in the diagrammatic series for the full vertex.

Next, we have the Kernel-1 diagrams that only depend on one bosonic frequency
and are depicted in \cfg{fig:k1}. Here, two pairs of (incoming or outgoing) lines enter at the respective same interaction $U$.
In this case the vertex depends only on  the total transferred frequency 
at these interactions (it is the same bosonic frequency 
for both pairs because  of energy conservation).\cite{Wentzell,Rohringer}
There are three diagrams  in \cfg{fig:k1}
and hence there are three Kernel-1 contributions each of which depends on a single bosonic frequency.
Switching from Matsubara frequencies to imaginary times, as defined in \ceq{eq:back_transform}, 
it turns out that the dependence on frequency differences corresponds to diagrams with
pairwise equal times.  That is, the diagrams shown in \cfg{fig:k1} correspond 
to the summation of all terms with two equal-time pairs.

For the Kernel-2 diagrams of  \cfg{fig:k2}, we have only
one pair of external legs that enter at the same $U$. Hence such diagrams depend on the transferred bosonic frequency at this $U$ and (because of energy conservation) 
one additional fermionic frequency of the unpaired legs.
This corresponds to one equal-time pair in Fourier space.
All the diagrams \cfg{fig:k1} and  \cfg{fig:k2} are two-particle reducible and thus, the asymptotic form of the full vertex
$F$ consists, apart from the constant background $U$, only of reducible terms $\Phi^{\ell}_{asympt}$.

\begin{figure}
\includegraphics{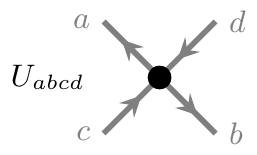}
\caption{\label{fig:umatrix} Diagram of the bare local interaction $U$. The bare vertex does not contain any Green's function, the (amputated) legs
  drawn in gray indicate the direction of the incoming/outgoing particles and their spin-orbital flavor $a,b,c,d$ (the Matsubara frequencies are suppressed for simplicity).}
\end{figure}

\begin{figure}
\includegraphics{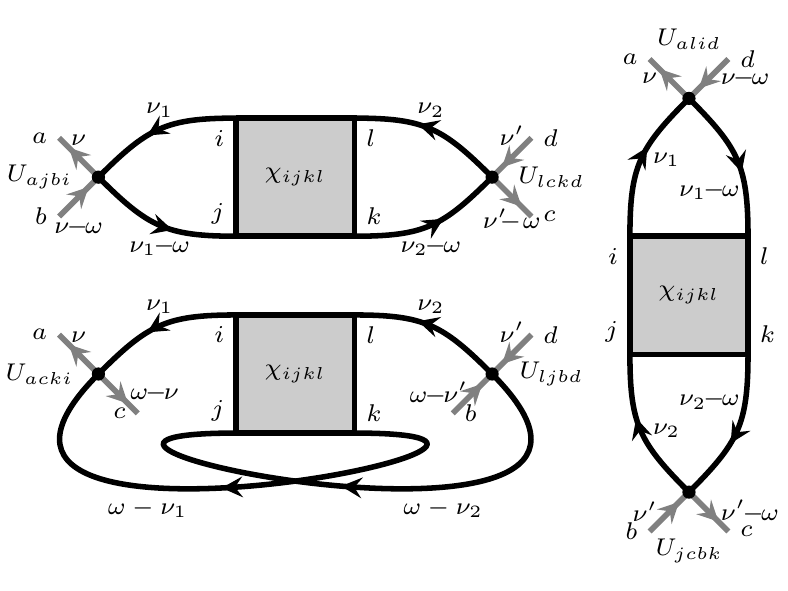}
\caption{\label{fig:k1} Vertex diagrams that depend on only one bosonic frequency, in $ph$-channel (top left), $\overbar{ph}$-channel (right)
  and $pp$-channel (bottom left). Frequencies are given in the channel-specific notation (see Appendix \ref{sec:freq-not}).}
\end{figure}

\begin{figure}
\includegraphics{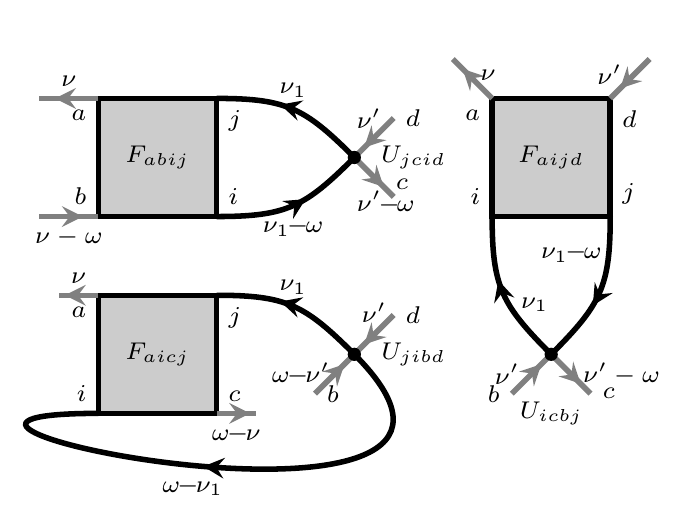}
\caption{\label{fig:k2} Vertex diagrams that depend on one bosonic and one fermionic frequency, in $ph$-channel (top left), $\overbar{ph}$-channel (right)
  and $pp$-channel (bottom left). Frequencies are given in the channel-specific notation (see Appendix \ref{sec:freq-not}).}
\end{figure}

\subsection{Equal-time two-particle Green's functions}\label{sec:Susc}

We now have to find a way to extract the aforementioned asymptotics from Green's function-like quantities, which
are accessible in impurity solvers such as CT-QMC. 

Considering the full Green's function $G_{ijkl}(\tau_1,\tau_2,\tau_3,\tau_4)$, we need to form two equal-time pairs for the diagrams of  \cfg{fig:k1} to arrive at
a function of two time arguments or one frequency-difference. There are three distinct ways to achieve this:
\begin{align}
\tau_1 = \tau_2 \equiv \tau,\; \tau_3 = \tau_4 \equiv  \tau' \label{eq:et_1} \\
\tau_1 = \tau_3 \equiv \tau,\; \tau_2 = \tau_4 \equiv  \tau' \label{eq:et_3} \\
\tau_1 = \tau_4 \equiv \tau,\; \tau_2 = \tau_3 \equiv  \tau' \label{eq:et_2}
\end{align}
which relate to the $ph$, $pp$ and $\overbar{ph}$ channel.
The ``two-legged'' two-particle Green's function for the $ph$-channel, defined in \eqref{eq:et_1}, is
\begin{multline}
\label{eq:p2phlg}
 G^{ph,\nu_1-\nu_2}_{ijkl} = \int \mathrm{d} \tau \mathrm{d} \tau' e^{i (\nu_1-\nu_2) (\tau - \tau')} \times \\
 \langle T_\tau d_i(\tau) d^\dagger_j(\tau) d_k(\tau') d^\dagger_l(\tau') \rangle,
\end{multline}

and for the $pp$-channel, we get
\begin{multline}
\label{eq:p2pp}
 G^{pp,\nu_1+\nu_3}_{ijkl} = \int \mathrm{d} \tau \mathrm{d} \tau' e^{i (\nu_1 + \nu_3) (\tau - \tau')}  \times \\
 \langle T_\tau d_i(\tau) d^\dagger_j(\tau') d_k(\tau) d^\dagger_l(\tau') \rangle.
\end{multline}

While the above functions have to be measured separately, the third, related to the $\overbar{ph}$-channel, 
can be obtained from the first by the crossing relation (see Ref.~\onlinecite{GallerDGA} for an illustration)
\begin{equation}
\label{eq:crossing}
 G^{\overbar{ph}}_{ijkl} = -G^{ph}_{ilkj}
\end{equation}
and depends on the frequency difference $\nu_1-\nu_4$. 

From the six ways to form one equal-time pair as needed for the diagrams \cfg{fig:k2}, it is sufficient to
consider only the following three, with the others related by time-reversal
symmetry:
\begin{align}
\tau_1 \equiv \tau,\; \tau_2 \equiv \tau',\; \tau_3 = \tau_4 \equiv \tau'' \label{eq:et_4}, \\
\tau_1 \equiv \tau,\; \tau_3 \equiv \tau',\; \tau_2 = \tau_4 \equiv \tau'' \label{eq:et_6}, \\
\tau_1 \equiv \tau,\; \tau_4 \equiv \tau',\; \tau_2 = \tau_3 \equiv \tau''. \label{eq:et_5}
\end{align}
Here, Eqs.~\eqref{eq:et_4}-\eqref{eq:et_5} are related, as before, to the $ph$, $pp$ and $\overbar{ph}$ channel.
The ``three-legged'' two-particle Green's function in the $ph$-channel corresponding to  Eq.~\eqref{eq:et_4} follows as
\begin{multline}
\label{eq:p3phlg}
 G^{ph,\nu_1,\nu_1-\nu_2}_{ijkl} = \int \!\! \mathrm{d} \tau \mathrm{d} \tau' \mathrm{d} \tau'' 
 e^{i (\nu_1 (\tau-\tau') + (\nu_1-\nu_2) (\tau'\!- \tau''))} \times \\ 
 \langle T_\tau d_i(\tau) d^\dagger_j(\tau') d_k(\tau'') d^\dagger_l(\tau'') \rangle,
\end{multline}
and in the $pp$-channel (\ceq{eq:et_6}) it is

\begin{multline}
\label{eq:p3pp}
 G^{pp,\nu_1,\nu_1+\nu_3}_{ijkl} = \int \!\! \mathrm{d} \tau \mathrm{d} \tau' \mathrm{d} \tau'' 
 e^{i (\nu_1 (\tau - \tau') + (\nu_1+\nu_3) (\tau'\!- \tau''))} \times \\
 \langle T_\tau d_i(\tau) d^\dagger_j(\tau'') d_k(\tau') d^\dagger_l(\tau'') \rangle.
\end{multline}
Again, the Green's function in the $\overbar{ph}$-channel can be obtained by the crossing relation \ceq{eq:crossing},
the frequency arguments are then $\nu_1$ and $\nu_1-\nu_4$. Please note that 
$\nu_1-\nu_2$, $\nu_1+\nu_3$ and $\nu_1-\nu_4$ are referred to as the channel-specific 
bosonic Matsubara frequencies $\omega_{ph}$, $\omega_{pp}$ and $\omega_{\overbar{ph}}$, respectively.
A full table with channel-specific frequency notations is given in Appendix \ref{sec:freq-not}.

\subsection{Subtraction of disconnected parts} 
We have seen in \ceq{eq:2green} and \cfg{fig:g2chi}, that the full two-particle Green's function,
as measured in CT-QMC, contains one connected and also two disconnected parts. Hence, in order to
arrive at the two- and three-legged diagrams of \cfg{fig:k1} and \cfg{fig:k2}, it is necessary
to eliminate the disconnected terms.
In the following we will assume the one-particle Green's function to be flavor diagonal, 
such that $ G_{ij}(\tau_1,\tau_2) \equiv G_i(\tau_1,\tau_2)\delta_{ij}$.
We recover the physical single-frequency susceptibility in the particle-hole channel by
subtracting the constant ``straight term'',
\begin{multline}
\label{eq:dc_phlg}
\chi_{ijkl}^{ph,\omega}=G_{ijkl}^{ph,\omega}
  - (1-n_i) (1-n_k) \delta_{\omega 0}\delta_{ij}\delta_{kl},
\end{multline}

whereas the particle-particle susceptibility is already given by
\begin{equation}
\label{eq:dc_pp}
\chi_{ijkl}^{pp,\omega}=G_{ijkl}^{pp,\omega}.
\end{equation}

We will now turn to the three-legged  Green's functions, where we are again interested only in the connected part corresponding to  \cfg{fig:k2}.
For the particle-hole channel we find
\begin{multline}
\label{eq:dc3_ph}
 \chi^{\text{c},ph,\nu\omega}_{ijkl} = G^{ph,\nu\omega}_{ijkl}
   - G_i^\nu \times \\ 
   \left[
     (n_k-1) \delta_{ij}\delta_{kl} \delta_{\omega 0} - G_k^{\nu-\omega}\delta_{il}\delta_{jk}
   \right]
\end{multline}

and for the particle-particle channel
\begin{multline}
\label{eq:dc3_pp}
 \chi^{\text{c},pp,\nu\omega}_{ijkl} = G^{pp,\nu\omega}_{ijkl}
   - \left( \delta_{ij}\delta_{kl} - \delta_{il}\delta_{jk} \right) G_i^\nu G_k^{\omega-\nu}.
\end{multline}
As usual, the corresponding expressions for the transverse particle-hole channel can be 
obtained by applying the crossing relation \ceq{eq:crossing}.

\subsection{Kernel functions}\label{sec:Kernels}

After the subtraction of the disconnected parts from the two-particle Green functions,
the next step is to contract the equal-time legs with interaction vertices.
The two-legged objects have two pairs of equal times and therefore need two distinct bare vertices
to contract their legs and obtain the Kernel-1 functions $K^{(1),\ell}$:
\begin{align}
K^{(1),ph,\omega}_{abcd} &= -\sum_{ijkl} U_{ajbi} \, \chi_{ijkl}^{ph,\omega} \, U_{lckd}\label{eq:K1_1}\\
K^{(1),\overbar{ph},\omega}_{abcd} &= -\sum_{ijkl} U_{alid} \, \chi_{ijkl}^{\overbar{ph},\omega} \, U_{jcbk}\label{eq:K1_2}\\
K^{(1),pp,\omega}_{abcd} &= -\sum_{ijkl} \frac{U_{acki}}{2} \, \chi_{ijkl}^{pp,\omega} \, \frac{U_{ljbd}}{2}\label{eq:K1_3}
\end{align}
This corresponds precisely to the diagrams shown in~\cfg{fig:k1}.

For the Kernel-2 approximations, the procedure is a bit more involved.
After the bare vertex contraction, we need to amputate the remaining legs.
Thus, the Kernel-2 functions $K^{(2),\ell}$ in all three channels are
\begin{align}
K^{(2),ph,\nu\omega}_{abcd} &= \sum_{ij} \frac{-\chi^{\text{c},ph,\nu\omega}_{abji}}{G_a^{\nu}G_b^{\nu-\omega} } \, U_{icjd}
  -  K^{(1),ph,\omega}_{abcd}\label{eq:K2_1}\\
K^{(2),\overbar{ph},\nu\omega}_{abcd} &= \sum_{ij} \frac{-\chi^{\text{c},\overbar{ph},\nu\omega}_{aijd}}{G_a^{\nu}G_d^{\nu-\omega}} \, U_{icbj}
  -  K^{(1),\overbar{ph},\omega}_{abcd}\label{eq:K2_2}\\
K^{(2),pp,\nu\omega}_{abcd} &= \sum_{ij} \frac{-\chi^{\text{c},pp,\nu\omega}_{aicj}}{G_a^{\nu}G_c^{\nu-\omega}} \, \frac{U_{jibd}}{2}
  -  K^{(1),pp,\omega}_{abcd},\label{eq:K2_3}
\end{align}
where we had to subtract the Kernel-1 functions in order to avoid double-counting of diagrams.

Now we have six functions going to zero for high frequencies $\nu$ or $\omega$, from which we can compile
the asymptotic vertex. 

\subsection{Asymptotic form of the full vertex}

According to the (local) parquet equation, the full vertex $F_{abcd}$ can be decomposed into a fully irreducible and several reducible parts:
\begin{equation}
F_{abcd}^{\nu\nu'\omega} = \Lambda_{abcd}^{\nu\nu'\omega} + \Phi_{abcd}^{ph,\nu\nu'\omega} + \Phi_{abcd}^{\overbar{ph},\nu\nu'\omega} + \Phi_{abcd}^{pp,\nu\nu'\omega}.
\end{equation}
We are now able to construct the asymptotic form of the reducible vertices $\Phi$ using:\cite{Wentzell}
\begin{equation}
\Phi^{\text{asympt},\ell,\nu\nu'\omega}_{abcd} = K^{(1),\ell,\omega}_{abcd} + K^{(2),\ell,\nu\omega}_{abcd} + \overbar{K}^{(2),\ell,\nu'\omega}_{abcd},
\end{equation}
where the functions $\overbar{K}^{(2),\ell}$ are found to be equal to $K^{(2),\ell}$ due to time-reversal symmetry.
Therefore summing up all $K^{(i),\ell}$, we get the asymptotic form of the full vertex:
\begin{multline}
\label{eq:F-asympt}
F^{\text{asympt}}_{abcd}(\nu_\ell,\nu_\ell',\omega_\ell)  - U_{abcd} =\\
  K^{(1),ph,\omega_{ph}}_{abcd} 
    + K^{(2),ph,\nu_{ph}\omega_{ph}}_{abcd}
    + K^{(2),ph,\nu'_{ph}\omega_{ph}}_{abcd}\\
  + K^{(1),\overbar{ph},\omega_{\overbar{ph}}}_{abcd}
    + K^{(2),\overbar{ph},\nu_{\overbar{ph}}\omega_{\overbar{ph}}}_{abcd} 
    + K^{(2),\overbar{ph},\nu'_{\overbar{ph}}\omega_{\overbar{ph}}}_{abcd}\\
  + K^{(1),pp,\omega_{pp}}_{abcd}
    + K^{(2),pp,\nu_{pp}\omega_{pp}}_{abcd}
    + K^{(2),pp,\nu'_{pp}\omega_{pp}}_{abcd}
\end{multline}
In this way we are now able to build arbitrarily large vertices in any frequency notation,
which leads to significant improvements of further calculations. 

\section{Implementation}\label{sec:Worm}

\subsection{Worm Sampling}
For the calculation of the equal-time two-particle Green's functions we employ the hybridization expansion (CT-HYB)~\cite{Werner} due to its favorable scaling
at finite temperature and its ability to treat general local interactions efficiently. The traditional formulation of the CT-HYB algorithm assumes importance
sampling and explores the phase space of the partition function $Z$. 
One- and two-particle Green's function are then obtained by ``removing'' hybridization lines.
For non-density-density interactions, this is in general not possible. Instead, we hence use a worm algorithm recently introduced to CT-HYB,~\cite{Gunacker,Gunacker16} and
measure  equal-time  two-particle correlation functions, which are then transformed into the kernel functions (\ref{eq:K1_1}) - (\ref{eq:K2_3}) in a post-processing step.

Worm sampling stands in contrast to partition function sampling as we no longer explore the phase space $\mathcal{C}_Z$ of the partition function, but rather an
extended phase space $\mathcal{C}_W$ for an
extended partition function ${W=Z+ \eta Z_G}$, where $Z_G$ is the partition function of an exemplary worm space and $\eta$ the relative balancing factor. While we
sample configurations, which do not represent the
denominator of the expectation value, we profit due to more flexibility in defining the estimator. The exact procedure on how to define equal-time Green's function
estimators can be found in previous works.~\cite{Gunacker16} By adding the local creation and annihilation operators of the estimators to the local trace of the
infinite perturbation series in the hybridization expansion, one effectively switches to worm space. We redefine the single-frequency expectation values in
Eqs.~(\ref{eq:p2phlg})-(\ref{eq:p2pp}) in terms of worm estimators:
\begin{equation}
\label{eq:p2mc}
 G^{\ell,\omega}_{\mathcal{C}_{1,\ell}} = \langle \mathrm{sgn}\times e^{i\omega(\tau-\tau')} \rangle_{MC},
\end{equation}
where $\mathcal{C}_{1,\ell}$ are the configuration spaces of the particle-hole and particle-particle single-frequency estimator and `sgn' denotes the
sign of the configuration.
Further, the two-frequency expectation values in Eqs.~(\ref{eq:p3phlg})-(\ref{eq:p3pp}) follow as:
\begin{equation}
\label{eq:p3mc}
 G^{\ell,\nu\omega}_{\mathcal{C}_{2,\ell}} = \langle \mathrm{sgn} \times e^{i(\nu(\tau-\tau')+\omega(\tau'-\tau''))} \rangle_{MC},
\end{equation}
where $\mathcal{C}_{2,\ell}$ are the configuration spaces of the particle-hole and particle-particle two-frequency estimator. 
We emphasize that the measured quantities still need to be normalized with respect to the partition function.

Apart from the above estimators assuming $\delta$-like bins, we have further implemented estimators considering the entire configuration as suggested in Ref.
\onlinecite{Shinaoka}.  At this point we note that for density-density interactions the worm algorithm is not necessary. Instead an implementation of the 
estimators in a segment algorithm is more feasible. In another context, the three-legged estimator was already defined for the segment representation.
\cite{Hafermann_ret}

\subsection{Numerical Effort}
\label{sec:numerical_effort}
In terms of the numerical effort of calculating the vertex asymptotics we benefit twofold. Firstly, the asymptotics scale quadratically in the number of frequencies $\sim \#w^2$, whereas the calculation
of the full two-particle object scales cubically  $\sim \#w^3$. In the asymptotical region, the three dimensional Fourier transform is thus replaced by a two dimensional
transform. By sampling a two-dimensional phase space instead of a three-dimensional one, we effectively collect more data-points for each imaginary time bin
which reduces the noise. Secondly, the non-asymptotic region needs to be calculated on a much smaller grid, that is, the prefactor of the full vertex measurement is
greatly reduced. Besides saving computational time, calculating the asymptotics also saves storage which for $M$-orbital vertices 
is  $\sim \#w^3 M^4$, so that storing the vertex easily requires  Giga- and Tera-Bytes. 

Due to the parameterization of the vertex function we can introduce cut-offs, as already suggested elsewhere.\cite{Kunes} While this  effect is hardly captured in
terms of numerical efficiency, this allows us to extend the asymptotic structure to arbitrary box sizes. As a  consequence, box summations do not suffer from finite size box effects.
\subsection{Workflow}
Having explained the calculation of Green's functions in QMC, we consider it useful to summarize
the whole workflow at this point:
\begin{enumerate}
\item QMC-calculation of $G_i^\nu$, $G_{ijkl}^{\nu\nu'\omega}$, $G_{ijkl}^{ph/pp,\omega}$,
    $G_{ijkl}^{ph/pp,\nu\omega}$, $n_i$ [Eqs.~(\ref{eq:p2phlg}), (\ref{eq:p2pp}), (\ref{eq:p3phlg}), (\ref{eq:p3pp})],
\item Subtraction of disconnected terms to obtain susceptibilities $\chi_{ijkl}^{ph/pp,\omega}$ 
    and connected diagrams $\chi_{ijkl}^{\text{c},ph/pp,\nu\omega}$ [Eqs.~(\ref{eq:dc_phlg})-(\ref{eq:dc3_pp})],
\item Amputation of legs from $\chi_{ijkl}^{\text{c},ph/pp,\nu\omega}$ [contained in Eqs.~(\ref{eq:K2_1})-(\ref{eq:K2_3})],
\item U-matrix contractions [Eqs.~(\ref{eq:K1_1})-(\ref{eq:K2_3}) $\rightarrow$ Kernel-1 functions ready at this point],
\item Subtraction of Kernel-1 functions from the connected diagrams in order to get the Kernel-2 functions 
    [contained in Eqs.~(\ref{eq:K2_1})-(\ref{eq:K2_3})].
\item Construction of $F^{\text{asympt}}$ from the Kernel functions [\ceq{eq:F-asympt}],
\item Combination of full $F$ and $F^{\text{asympt}}$.
\end{enumerate}
We note, however, that it is recommendable for most applications to store only the Kernel functions permanently,
and construct the asymptotically extended vertex ``on the fly'' during a calculation in which it is used. 

\section{Results}\label{sec:Results}

\subsection{Single-orbital Hubbard model}
The Hubbard model is an often employed model for strongly correlated electrons 
on a lattice. Its Hamiltonian consists of a hopping term, capturing the kinetic 
energy of the electrons, and a local interaction term that models their on-site Coulomb
repulsion. Formally, the kinetic term is related to a tight-binding model, and
the local interaction has the same form as for the AIM  \ceq{eq:anderson}. For the single-orbital case with next-neighbor
hopping only, the Hamiltonian reads 
\begin{equation}
\label{eq:hubbard-oneband}
H=-t\sum_{\langle i,j \rangle,\sigma} c_{i\sigma}^\dag c_{j\sigma} + U \sum_i c_{i\uparrow}^\dag c_{i\downarrow}^\dag c_{i\downarrow} c_{i\uparrow},
\end{equation}
where $t$ is the hopping parameter and $U$ the Hubbard interaction. Indices $i$ and $j$ denote lattice sites here, and $\sigma$ stands for
the spin projection. \\
For a three-dimensional simple cubic lattice, the hopping term 
determines the bandwidth of the system as $W=12t$ and the standard deviation as $D/2=\sqrt{6}t$.\cite{SchaeferQCP,Rohringer2011}
Thereafter, all energies concerning the Hubbard model will be measured in units of $D\equiv 1$. 
The model studied here is characterized by an interaction strength of $U=2D$
at an inverse temperature of $\beta=8/D$. \\
DMFT~\cite{Metzner,Georges:1992,Kotliar_dmft,Held} provides a possibility to solve the Hubbard
model in the limit of infinite dimensions by self-consistently mapping it onto
an auxiliary AIM. In finite dimensions, this corresponds to approximating
the self-energy to be purely local. There exists a variety of solvers for
the impurity problem; we employ CT-QMC using the \textsc{w2dynamics}  package.\cite{Parragh,Wallerberger}

Whereas QMC in principle provides a numerically exact solution, it suffers
from statistical uncertainty, making it reasonable to benchmark
against exact diagonalization (ED).\cite{Caffarel} To this end, we solve  by QMC the impurity problem specified by the bath
parameters of a converged ED calculation.\cite{SchaeferQCP}

In order to give an overall impression of the situation, we show a slice
of the full vertex $F$ in \cfg{fig:f_hub}. The spin-components 
$F_\uu\equiv F_\uuuu$, $F_\ud\equiv F_\uudd$, and $F_\udb\equiv F_\uddu$\cite{Rohringer}
were combined to the density and magnetic channel by
\begin{align}
F_{d}&=F_{\uu}+F_{\ud}\\
F_{m}&=F_{\uu}-F_{\ud}\overset{\text{SU(2)}}{=}F_{\udb}.
\end{align}
The first column shows vertices calculated by the improved-estimator method with
worm sampling in about 30000 CPU hours. In the second column, the data
in the asymptotic regions, defined by
\begin{equation}
\label{eq:replacement-condition}
\nu_1\nu_2\nu_3\nu_4 \frac{\beta^4}{\pi^4} > l^4\left|\delta_{\nu_1\nu_2}+\delta_{\nu_1\nu_4}-\delta_{\nu_1\nu_2}\delta_{\nu_1\nu_4}\right|^4,
\end{equation}
were replaced according to the method proposed in this article, 
with a replacement parameter of $l=10$.
For comparison, we show ED results in the third column. The replacement procedure \ceq{eq:replacement-condition} is motivated by atomic limit calculations in Appendix \ref{sec:atomic}. 
\begin{figure}
\includegraphics{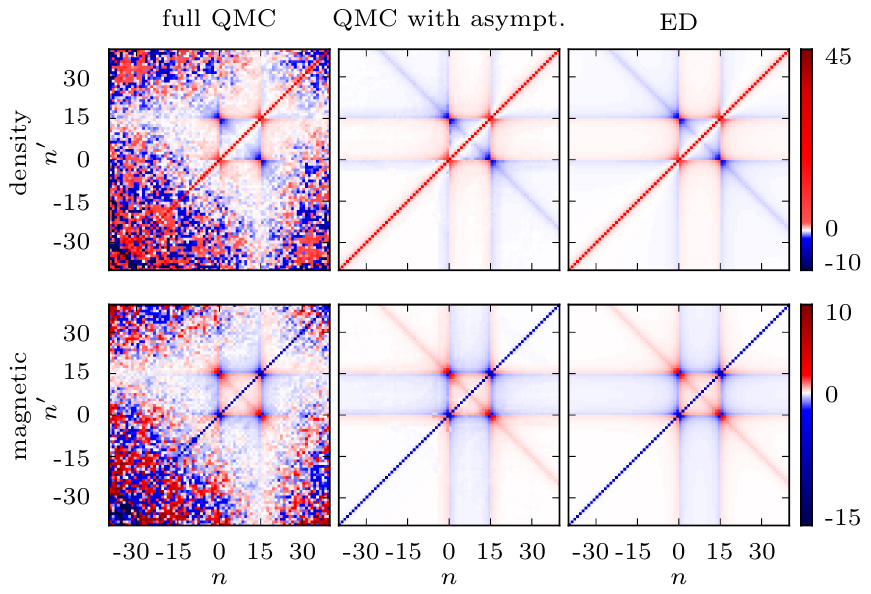}
\caption{\label{fig:f_hub}  (Color online) Local full vertex $F$ for the half-filled Hubbard model in DMFT for $U=2D$ and $\beta=8/D$. 
Upper row:  $F_{d}$ in the the density channel.
Lower row:  $F_m$ in the magnetic channel.
  First column: $F$ extracted from an improved-estimator CT-QMC measurement with full frequency dependence. 
  Second column: $F$ from QMC, combined with asymptotics according to \ceq{eq:replacement-condition} with $l=10$.
  Third column:  $F$ obtained with ED for comparison.
 We use  the particle-hole frequency representation (see Appendix \ref{sec:freq-not}) and fix $\omega_{ph}=15\frac{2\pi}{\beta}$.
  }
\end{figure}

The statistical uncertainty of one- and two-particle Green's functions is in principle well controlled by the $1/\sqrt{N}$ scaling of the Monte Carlo method.
The amputation of four outer legs, however, corresponds to the division by four inverse one-particle Green’s functions, each asymptotically approaching zero.
Eventually, this leads to a strong amplification of noise in the full vertex function $F$ (first column of \cfg{fig:f_hub}).
The equal-time two-particle Green's functions, on the other hand, can be measured more accurately due to their reduced time (frequency) dependence. 
In order to calculate the Kernel-2 functions, only two legs need to be amputated, also resulting in a lower noise level (second column of \cfg{fig:f_hub}).

We observe a good qualitative agreement of the asymptotically improved vertex with ED (third column of \cfg{fig:f_hub}).
A more quantitative comparison
can be made by directly investigating the difference of the full vertex $F$ and its purely 
asymptotic version. This is shown in \cfg{fig:r-hub}, again in the density and magnetic channels, 
for three different values of the bosonic frequency $\omega_{ph}$ in the particle-hole channel. 
In good accordance to the theoretical foundation of the kernel functions, the magnitude of the
difference decreases for high values of any frequency. \\
\begin{figure}
\includegraphics{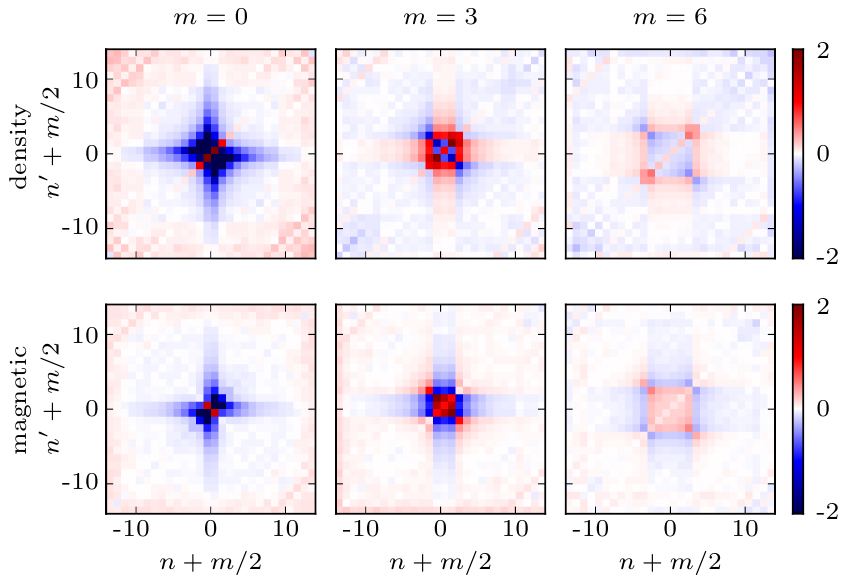}
\caption{\label{fig:r-hub} (Color online) Difference of ED vertex and asymptotic vertex in density (top) and magnetic channel  (bottom)
  for three different bosonic frequencies $\omega_{ph}=m\frac{2\pi}{\beta}$ (columns). The fermionic Matsubara frequencies on the $x$- and $y$-axis are shifted by $m$ as indicated.}
\end{figure}

To demonstrate the practical applicability of the vertex asymptotics,
one can calculate, for example, physical susceptibilities
\begin{equation}
\chi^{ph,\omega}_{d/m} = \frac{1}{\beta^2} \sum_{\nu\nu'} \chi^{ph,\nu\nu'\omega}_{d/m}.
\label{eq:suscsum}
\end{equation}
This is a reasonable test, because the physical susceptibilities can be computed also 
directly from the one-frequency Green's functions measured in QMC via \ceq{eq:dc_phlg}.
In \cfg{fig:chi_w_hub} we observe two effects brought about by the
asymptotics method: The results obtained by summing over a large frequency box
is slightly smoothed (best visible in the inset). This reduction of noise can be understood by comparing
the first two columns in \cfg{fig:f_hub}, where using the asymptotics of the vertex decreases the noise. 

The second effect  of using the asymptotic vertex is even larger and was our original motivation: the reduction
of the ``finite-box effect'' that is visible primarily in the density channel (upper panel of  \cfg{fig:chi_w_hub}). For
high values of the bosonic frequency argument, the physical susceptibility should
go to zero, as it is the case when it is measured directly in continuous time (solid line). If
it is calculated however by summation over fermionic Matsubara frequencies, the inevitable truncation leads to a wrong asymptotic
behavior. The deviation can be reduced only by including a larger frequency box
into the summation, which is easily possible using the vertex asymptotics. In principle there is no restriction to the box size here,
but we find it sufficient to sum over 1600$\times$1600 elements per bosonic
frequency, which would already be infeasible without asymptotics. 

\begin{figure}
\includegraphics[width=0.49\textwidth]{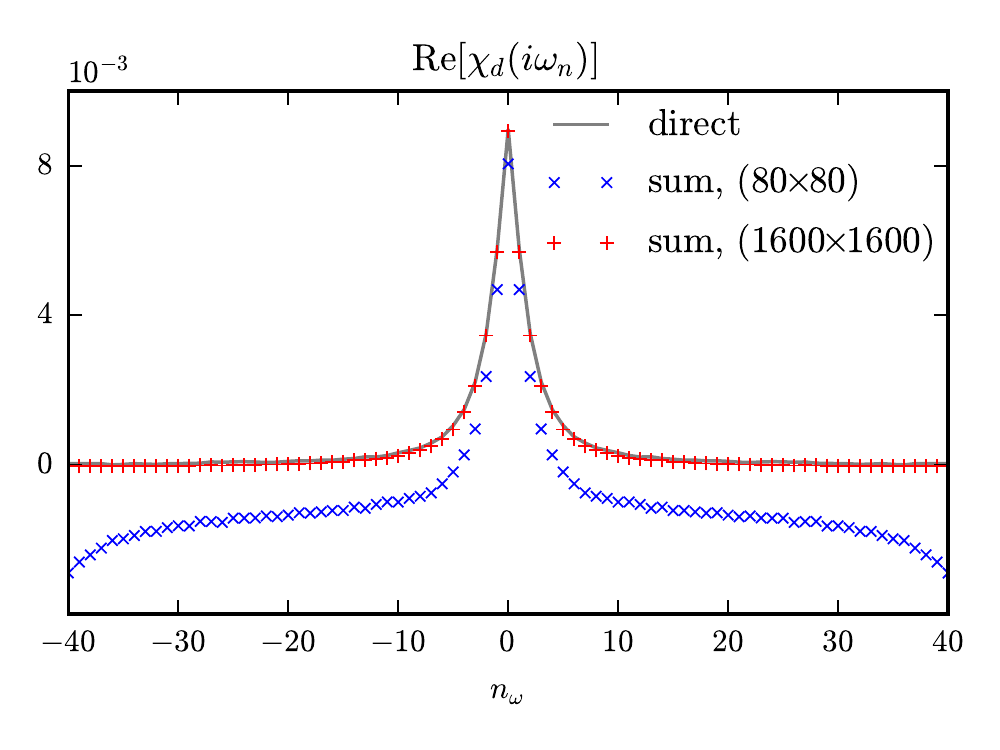}\\
\includegraphics[width=0.49\textwidth]{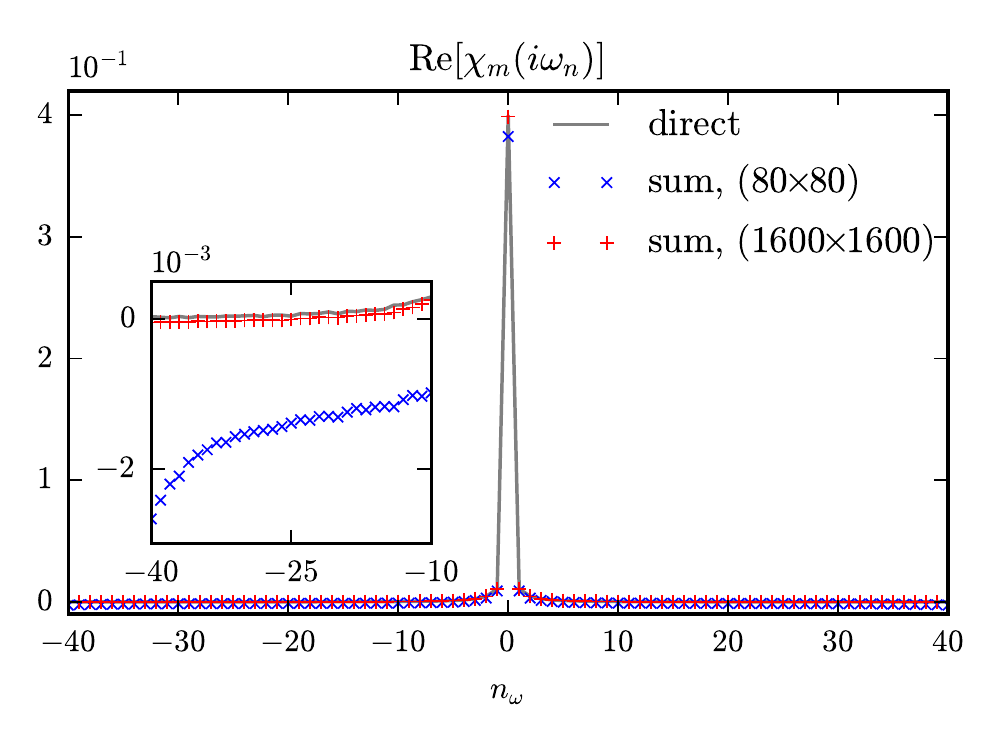}
\caption{\label{fig:chi_w_hub} Local susceptibilities $\chi_{\mathrm{loc}}(i\omega_n)$ in density (top) and magnetic channel (bottom) for the Hubbard model in DMFT at $U=2D$ and $\beta=8/D$.
The bosonic Matsubara frequencies are $\omega_n=n\frac{2\pi}{\beta}$. We compare the direct calculation via \ceq{eq:dc_phlg} (solid line) to that 
using the summation \ceq{eq:suscsum} over Fermionic Matsubara frequencies
without vertex asymptotics in a small box (x) to that using
vertex asymptotics and hence a large box (+). Inset: zoom in showing the box-effect and noise reduction.}
\end{figure}

\subsection{Multi-orbital test case: SrVO$_3$}
Since the derivations in the previous sections were done without restriction
to one-band models or density-density interaction, it is possible to apply
the procedure described above to a more general case. As a suitable material,
we  chose SrVO$_3$, which has a long tradition for benchmarking realistic material calculations using DMFT.\cite{Sekiyama2004,Ishida2006,Lee12,Taranto2013, PhysRevB.88.235110} Its band structure can be calculated by \textsc{wien2k},\cite{wien2k}
using the generalized gradient approximation. Subsequently, $t_{2g}$ bands,
 which cross the Fermi level, are projected onto maximally localized Wannier functions by 
\textsc{wien2wannier}.\cite{wien2wannier}
For these strongly correlated  $t_{2g}$ bands we consider a SU(2) symmetric Slater-Kanamori
interaction that is parameterized by an intra-orbital Hubbard $U$, an inter-orbital
$U'$ and Hund's coupling $J$. Calculations in constrained local density approximation
yield values of $U=5$eV, $J=0.75$eV and $U'=U-2J=3.5$eV.\cite{Sekiyama2004,Nekrasov}\\
The following DMFT calculation, as well as the calculation of the 
one-, two- and three-frequency two-particle Green's functions, was done by \textsc{w2dynamics}
at an inverse temperature of $\beta=10\mathrm{eV}^{-1}$.\\
Since we treat SrVO$_3$ as a three-orbital system, the two-particle objects have in general
have $(2\cdot3)^4=1296$ spin-orbital components, of which due to the structure of the interaction, 
however, only 126 are non-vanishing. If we use instead of all spin-components the density and
magnetic channels, which is possible for  SU(2) symmetry, the number of non-vanishing
components is reduced to 21 per channel. Furthermore the local vertex functions exhibit 
orbital symmetry that reduces the number of distinct components to 4 per channel in our case of degenerate orbitals. \\
In \cfg{fig:f_svo} a slice of the vertex with four equal band indices is shown in the density
and magnetic channel: $F_{d/m,1111}^{\nu\nu'\omega_{15}}$.
As before, in the left column we show the vertex, as calculated by amputation of external 
legs from the susceptibility with full frequency dependence. 
This is the way how the multi-orbital vertex was determined previously in AbinitioD$\Gamma$A calculations.\cite{GallerDGA}
In the right column,  we present the same vertex, but now
the data at asymptotic values of the frequency, given by \ceq{eq:replacement-condition} with $l=15$, 
are replaced by the asymptotic vertex. Our approach reduces the noise considerably and makes multi-orbital vertex calculations much more feasible. \\
\begin{figure}
\includegraphics{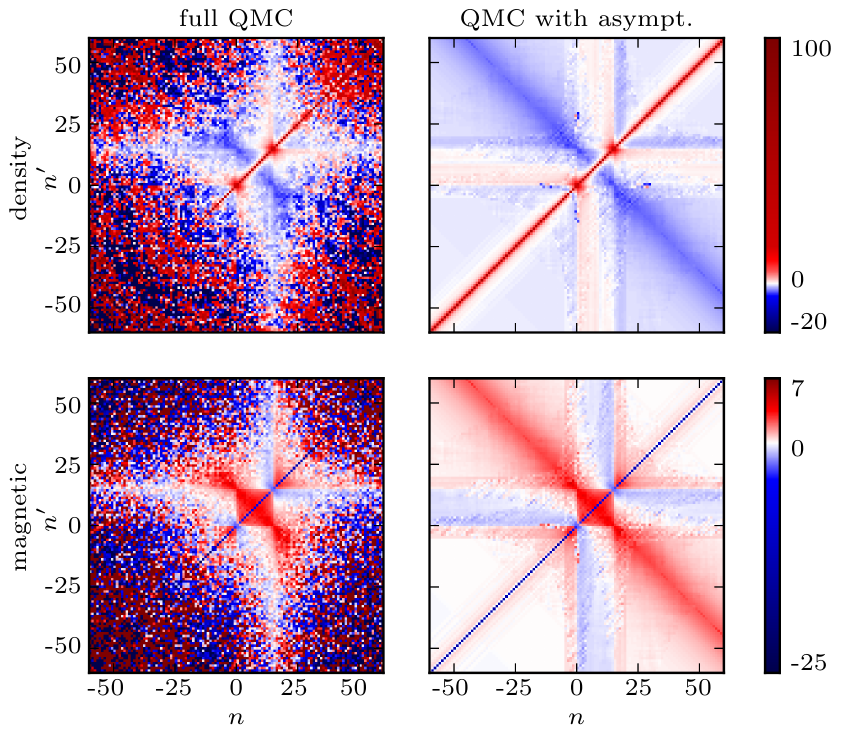}
\caption{\label{fig:f_svo} (Color online) Matrix element of the full vertex $F_{d,1111}^{\nu\nu'\omega_{15}}$ (upper row) 
  and $F_{m,1111}^{\nu\nu'\omega_{15}}$ (lower row) for four times at the same $t_{2g}$ orbital.
  Left column: $F$ extracted from an improved-estimator CT-QMC measurement with full frequency dependence. 
  Right column: $F$, combined with asymptotics according to \ceq{eq:replacement-condition} with $l=15$. 
  To remove the constant background, $F_d$ was shifted by $U_d=U$ and $F_m$  by $U_m=-U$.
  }
\end{figure}

In order to show  how the fully frequency-dependent vertex $F$   approaches  its
asymptotic form, we show in \cfg{fig:r-svo} three slices of the difference $F-F_{\mathrm{asympt}}$. Again a strong decay can be noticed, albeit slower
than in the Hubbard model studied above. Furthermore the diagonal defined by
$\nu_{ph}=\nu_{\overbar{ph}}$ is considerably more pronounced, a behavior
that is to be expected, however, by atomic limit calculations. \\
\begin{figure}
\includegraphics{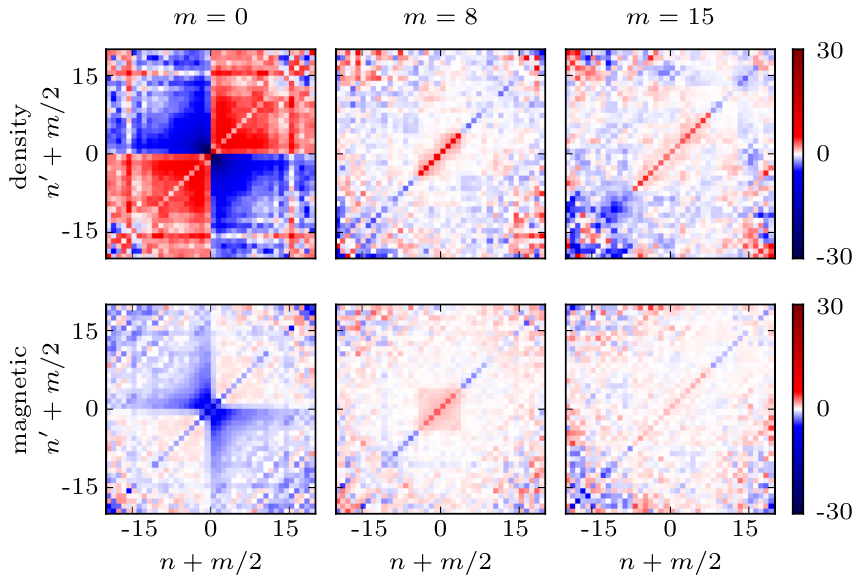}
\caption{\label{fig:r-svo} (Color online) Difference of $F_{d,1111}^{\nu\nu'\omega_{15}}$ (top)
  and $F_{m,1111}^{\nu\nu'\omega_{15}}$ (bottom) to their respective purely asymptotic version. 
  for different bosonic frequencies $\omega_m=\frac{2\pi}{\beta}m$.}
\end{figure}

A sample application of the asymptotics is again the calculation of 
frequency-summed susceptibilities. 
In order to demonstrate the ability of our method to treat pair hopping 
and spin flip contributions, introduced by the SU(2) symmetric Kanamori interaction,
we show the components $\chi^{d/m,\omega}_{1122}$ in \cfg{fig:chi_w_svo}.
Two important observations can be made in these plots: First,
the noise can be largely reduced in the high-frequency region, 
and second, large deviations at $\omega=0$ can be eliminated. 
\begin{figure}
\includegraphics[width=0.49\textwidth]{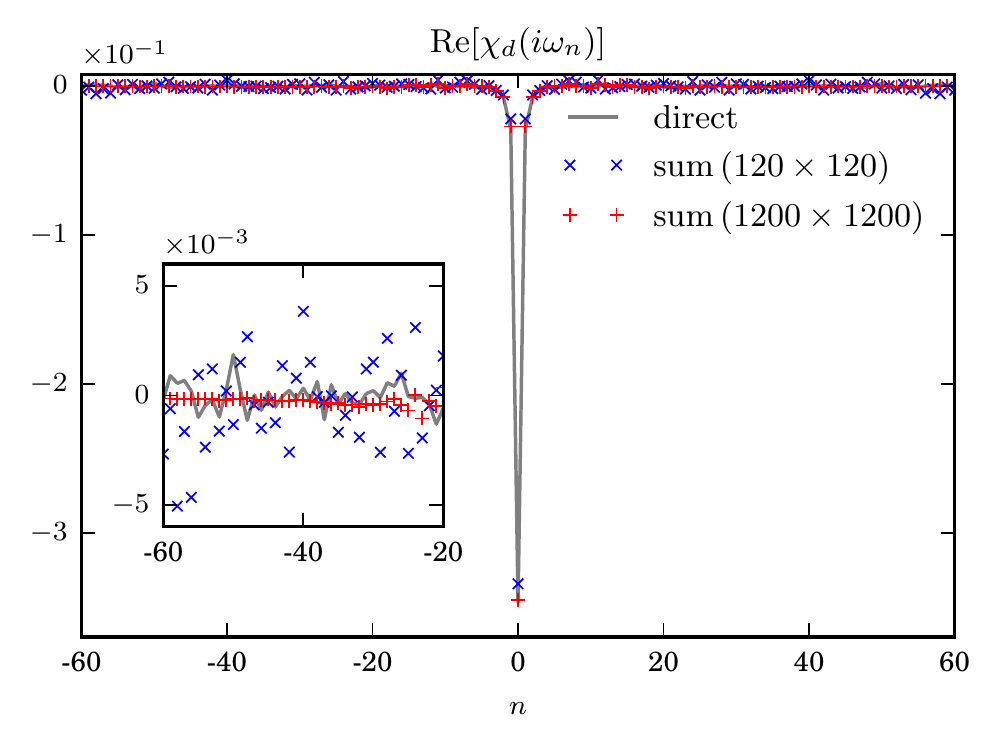}\\
\includegraphics[width=0.49\textwidth]{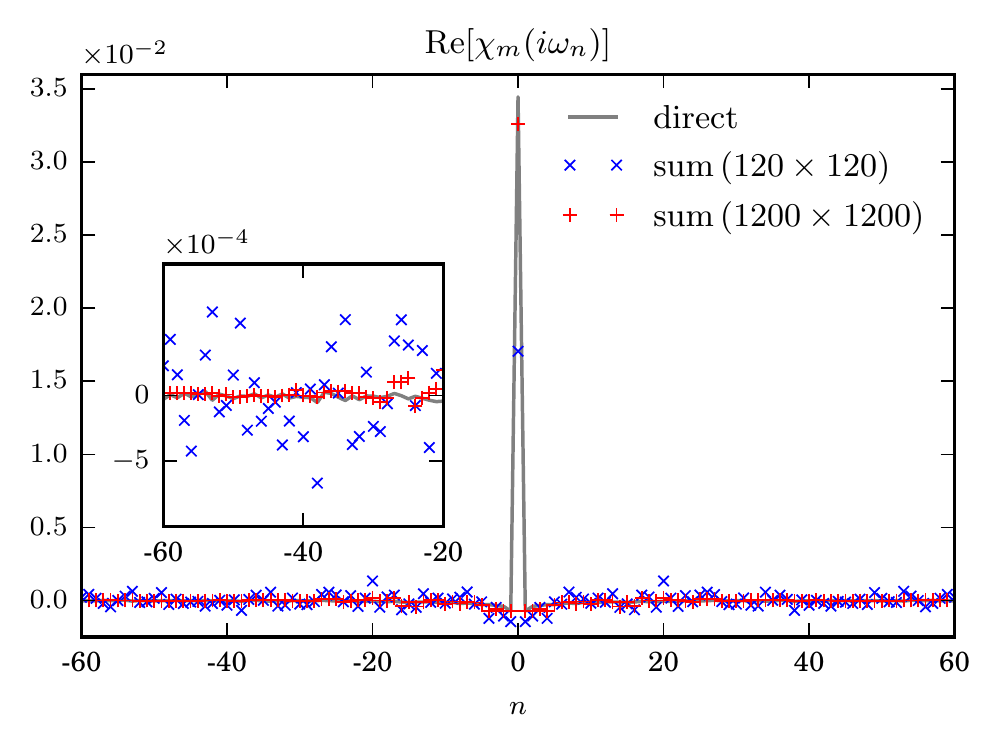}
\caption{\label{fig:chi_w_svo} (Color online) Local susceptibility $\chi^{\mathrm{loc}}_{d/m,1122}$ 
  of SrVO$_3$ between two different $t_{2g}$ orbitals in density (top) and magnetic channel (bottom). Inset: zoom in.}
\end{figure}


\section{Conclusion}\label{sec:Conclusion}
In this work we establish the link between reduced frequency (equal-time) two-particle Green's functions and the asymptotics of the  full vertex function $F$ for the multi-orbital AIM.
The former ones are, in principle, accessible by employing impurity solvers such as CT-QMC. We make use of a worm algorithm in the hybridization expansion to measure these equal-time Green's functions in CT-QMC for multiple orbitals and
general local interactions. From these Green's functions in turn,
we calculate the Kernel-1 and Kernel-2 functions for the vertex asymptotics. This requires  contractions with the bare interaction and a careful treatment of the disconnected parts. We benchmark the vertex asymptotics for the single-orbital Hubbard model in DMFT, by comparing our numerical CT-QMC data to ED results. As a second application, we calculate the vertex asymptotics for SrVO$_3$ using   three $t_{2g}$ orbitals for the low energy degrees of freedom.
In both cases, we demonstrate that using the asymptotics yields a much better
vertex with less noise and for an arbitrary large frequency box.
The latter allows us to avoid the errors  associated
a finite frequency box when calculating physical susceptibilities.

Our method allows us to assemble multi-orbital vertices in CT-QMC for arbitrary  frequency boxes, 
at a much reduced  computational time and storage. 
A second advantage is that we 
overcome the problem of noisy QMC vertices at larger frequencies.
Our paper  is hence a crucial step for making the (multi-orbital) vertex
available both for calculating general DMFT susceptibilities  and for diagrammatic extensions to DMFT.

\acknowledgments
We thank N. Wentzell, J. Kune\v{s}, A. Toschi, T. Ribic, D. Springer, A. Katanin, G. Li, and P. Thunstr\"om for valuable discussions.
In particular, we thank T. Sch\"afer for the ED reference data, and A. Galler for the SrVO$_3$ cooperation.
This work has been supported  by the  Vienna Scientific Cluster (VSC) Research Center funded by the Austrian Federal Ministry of Science, Research and Economy (bmwfw) and the European Research Council under the European Union's Seventh Framework Programme (FP/2007-2013) through ERC grant agreement n. 306447 (AbinitioD$\Gamma$A).  
The computational results presented have been achieved using the VSC. The plots were made using the matplotlib \cite{Hunter:2007}
 plotting library for \textsc{Python}.
 
\appendix

\section{Frequency mappings}\label{sec:freq-not}

Two-particle functions have four fermionic frequency arguments $\nu_1\ldots\nu_4$. 
Due to energy conservation,  one of the arguments is redundant and we use one bosonic
and two fermionic frequency arguments instead. 
Since the mapping between those two sets of frequencies is ambiguous, there exist
different possibilities that can be associated to the scattering channels.
They are called \emph{particle-particle} notation ($pp$), 
\emph{particle-hole} notation ($ph$) and \emph{transverse particle-hole} notation ($\overbar{ph}$). \\
We thus introduce, in addition to  $\nu_1\ldots\nu_4$, 
the \emph{particle-particle} frequencies $\nu_{pp}$, $\nu'_{pp}$ and $\omega_{pp}$; 
the \emph{particle-hole} frequencies $\nu_{ph}$, $\nu'_{ph}$ and $\omega_{ph}$;
and the \emph{transverse particle-hole} frequencies $\nu_{\overbar{ph}}$, $\nu'_{\overbar{ph}}$ and $\omega_{\overbar{ph}}$.
They are defined in the following way:

\begin{align}
\nu_1 &= \nu_{pp}              &=& \nu_{ph}               &=& \nu_{\overbar{ph}}\\
\nu_2 &= \omega_{pp}-\nu'_{pp} &=& \nu_{ph}-\omega_{ph}   &=& \nu'_{\overbar{ph}}\\
\nu_3 &= \omega_{pp}-\nu_{pp}  &=& \nu'_{ph}-\omega_{ph}  &=& \nu'_{\overbar{ph}}-\omega_{\overbar{ph}}\\
\nu_4 &= \nu'_{pp}             &=& \nu'_{ph}              &=& \nu_{\overbar{ph}}-\omega_{\overbar{ph}}
\end{align}

It is convenient to express all frequencies in all possible combinations:

\begin{align}
\nu_{pp}    &= \nu_{ph}                           = \nu_{\overbar{ph}} \\
\nu'_{pp}   &= \nu'_{ph}                          = \nu_{\overbar{ph}}-\omega_{\overbar{ph}} \\
\omega_{pp} &= \nu_{ph}+\nu'_{ph}-\omega_{ph}     = \nu_{\overbar{ph}}+\nu'_{\overbar{ph}}-\omega_{\overbar{ph}} \\
\omega_{ph} &= \nu_{pp} + \nu'_{pp} - \omega_{pp} = \nu_{\overbar{ph}} - \nu'_{\overbar{ph}} \\
\nu'_{\overbar{ph}} &= \nu_{ph}-\omega_{ph} = \omega_{pp}-\nu'_{pp} \\
\omega_{\overbar{ph}} &= \nu_{ph}-\nu'_{ph} = \nu_{pp}-\nu'_{pp}
\end{align} 

\section{Calculations in the atomic limit}\label{sec:atomic}
We also validated our approach in the atomic limit, which
is obtained by setting the hybridization to $V=0$ in the Anderson impurity model, i.~e.
\begin{equation}
  H = -\mu(n_\uparrow + n_\downarrow) + U n_\uparrow n_\downarrow.
\end{equation}
In this case, expectation values in the grand canonical ensemble with a Boltzmann weight
$\rho \sim\mathrm{exp}[-\beta H]$ and a chemical potential $\mu$ can be calculated analytically in the Lehmann basis 
$\left\lbrace \left|0\right>, \left| \uparrow \right>, \left| \downarrow \right>, \left| \uparrow\downarrow \right> \right\rbrace$.
At half filling, $\mu=U/2$ and thus, $\rho=\mathrm{diag}[1,e^{\beta\mu},e^{\beta\mu},1]/(2+2e^{\beta\mu})$.
Expectation values can be calculated as $\langle \mathcal{O} \rangle=\mathrm{Tr}[\mathcal{O}\rho]$.
In this way, one can calculate the full two-particle Green's function and, subsequently, the full vertex $F$.\cite{Rohringer,Georgthesis2013}
In Ref.~\onlinecite{Wentzell} the Kernel functions were calculated by taking high-frequency limits (see Eq.~(15) in Ref.~\onlinecite{Wentzell}).

On the other hand, we can obtain the vertex asymptotics via the procedure derived in Section \ref{sec:Derivation} of the present paper.
To this end, we first need to calculate the equal-time two-particle Green's functions, which are given in
Table \ref{tab:p2} and Table \ref{tab:p3}, using the Fermi function $f(\varepsilon) \equiv 1/(1+e^{\beta\varepsilon})$ as an abbreviation.

\begin{table}
\begin{center}
\bgroup
\def\arraystretch{2}
\begin{tabular}{l | c | c}
$G_{\sigma\sigma'}^{\ell,\omega}$ & $pp$ & $ph$ \\ \hline
$\uparrow\uparrow$ & 
  $0$ & 
  $\frac{\beta}{2}\delta_{\omega 0}$ \\
$\uparrow\downarrow$ & 
  $\frac{\beta}{2} f\left(\frac{U}{2}\right) \delta_{\omega 0}$ & 
  $\frac{\beta}{2} f\left(\frac{U}{2}\right) \delta_{\omega 0}$ \\
$\overbar{\uparrow\downarrow}$ & 
  $-\frac{\beta}{2} f\left(\frac{U}{2}\right) \delta_{\omega 0}$ & 
  $\frac{\beta}{2} f\left(-\frac{U}{2}\right) \delta_{\omega 0}$
\end{tabular}
\egroup
\end{center}
\caption{\label{tab:p2}Two-legged two-particle Green's functions in the atomic limit, i.e. \ceq{eq:p2phlg} and \ceq{eq:p2pp}, in particle-particle and particle-hole channel, respectively. 
  Frequencies are given in the channel-specific notations, see Appendix \ref{sec:freq-not}.}
\end{table}

\begin{table}
\begin{center}
\bgroup
\def\arraystretch{2}
\begin{tabular}{l | c }
$G_{\sigma\sigma'}^{pp,\nu\omega}$ &  \\ \hline
$\uparrow\uparrow$ &
  $0$ \\
$\uparrow\downarrow$ & 
  $\frac{\nu(\nu-\omega)-\frac{U^2}{4}}{\left(\nu^2+\frac{U^2}{4}\right)\left( (\nu-\omega)^2 + \frac{U^2}{4} \right)}
    -\delta_{\omega 0}\frac{\beta\frac{U}{2}f\left(\frac{U}{2}\right)}{\nu^2+\frac{U^2}{4}}$ \\
$\overbar{\uparrow\downarrow}$ & 
  $-\frac{\nu(\nu-\omega) - \frac{U^2}{4}}{\left(\nu^2+\frac{U^2}{4}\right)\left( (\nu-\omega)^2 + \frac{U^2}{4}\right)}
    +\delta_{\omega 0}\frac{\beta\frac{U}{2}f\left(\frac{U}{2}\right)}{\nu^2+\frac{U^2}{4}}$ \vspace{5mm}\\

$G_{\sigma\sigma'}^{ph,\nu\omega}$ &  \\ \hline
$\uparrow\uparrow$ &
  $\frac{\nu(\nu-\omega)-\frac{U^2}{4}}{\left(\nu^2+\frac{U^2}{4}\right)\left( (\nu-\omega)^2 + \frac{U^2}{4} \right)}
    + \delta_{\omega 0}\frac{\beta}{2} \frac{\frac{U}{2}\mathrm{tanh}\frac{\beta U}{4} + i\nu}{\nu^2+\frac{U^2}{4}}$ \\
$\uparrow\downarrow$ & 
  $ \delta_{\omega 0}\frac{\beta}{2} \frac{-\frac{U}{2} + i\nu}{\nu^2+\frac{U^2}{4}} $ \\
$\overbar{\uparrow\downarrow}$ & 
  $\frac{\nu(\nu-\omega)-\frac{U^2}{4}}{\left(\nu^2+\frac{U^2}{4}\right) \left( (\nu-\omega)^2 +\frac{U^2}{4}  \right)}
    +\delta_{\omega 0}\frac{\beta\frac{U}{2}f\left(-\frac{U}{2}\right)}{\nu^2+\frac{U^2}{4}}$
\end{tabular}
\egroup
\end{center}
\caption{\label{tab:p3}Three-time two-particle Green's functions in the atomic limit, i.e. \ceq{eq:p3phlg} and \ceq{eq:p3pp}, in particle-particle and particle-hole channel.
  Frequencies are given in the channel-specific notations, see Appendix \ref{sec:freq-not} }
\end{table}

In the following we will calculate only the $\ud$-components of the Kernel functions in the $ph$-channel explicitly,
but all components are given in Table \ref{tab:k1} and Table \ref{tab:k2}.
First, the single-frequency susceptibility is recovered from the respective Green's function by subtracting
the constant density term $\beta\delta_{\omega 0}/4$:
\begin{equation}
\chi^{ph,\omega}_\ud = G^{ph,\omega}_\ud - \frac{\beta\delta_{\omega 0}}{4} 
  = -\frac{1}{4}\beta\delta_{\omega 0}\left[ f\left(-\frac{U}{2}\right) - f\left(\frac{U}{2}\right)\right].
\end{equation}
Since the single-orbital U-matrix has only four non-vanishing components $U_\udud=U_\dudu = U$ and $U_\uddu=U_\duud = -U$, 
the Kernel function $K^{(1),ph,\omega}_\ud$ is directly related to $\chi^{ph,\omega}_\du$ by \eqref{eq:K1_1}:
\begin{equation}
\label{eq:k1-atomic}
K^{(1),ph,\omega}_\ud = - U^2 \chi^{ph,\omega}_\du
\end{equation}
Table  \ref{tab:k1} lists the other Kernel-I functions.

In order to extract $K^{(2),ph,\nu\omega}_\ud$ from equal-time two-particle Green's functions, 
it is of advantage to rewrite the latter, emphasizing their connection to one-particle Green's functions.
Since the U-matrix contraction relates $K^{(2),ph,\nu\omega}_\ud$ to $G^{ph,\nu\omega}_\uu$ only, we 
print the $\uu$-component:
\begin{multline}
G^{ph,\nu\omega}_\uu = -\frac{1}{2}\beta\delta_{\omega 0} G^\nu - G^\nu G^{\nu-\omega} + \\
  + \underbrace{
      \left[ \frac{U^2}{4\nu(\nu-\omega)} - \frac{K^{(1),ph,\omega}_\ud}{U}\left( 1+\frac{U^2}{4\nu^2} \right) \right]
    }_{\equiv -L^{ph,\nu\omega}_\uu}G^\nu G^{\nu-\omega}.
\end{multline}
From this, the kernel part $L^{ph,\nu\omega}_\uu$ is obtained by subtracting the disconnected parts (first line of the right-hand side) and 
amputating the legs $G^\nu G^{\nu-\omega}$. In a final step, the Kernel function $K^{(2),ph,\nu\omega}_{\ud}$
follows as
\begin{multline}
\label{eq:k2-atomic}
K^{(2),ph,\nu\omega}_{\ud}=U L^{ph,\nu\omega}_\uu - K^{(1),ph,\omega}_{\ud}\\
=\frac{U^2}{4\nu(\nu-\omega)}(K^{(1),pp,\omega}_\ud-U).
\end{multline}
Table \ref{tab:k2} lists the other Kernel-2 functions.
Apart from the different frequency conventions, our formulas agree with the results reported previously.\cite{Wentzell}

Using \eqref{eq:k1-atomic}, \eqref{eq:k2-atomic} and the crossing relation \eqref{eq:crossing} to 
calculate the Kernel functions in the $\overbar{ph}$-channel, we can now compile the full
asymptotic vertex from its $ph$-, $\overbar{ph}$- and $pp$-contributions. 
This is illustrated in \cfg{fig:vertex-assembly-atomic}, where each of the pictures corresponds
to one line of the right-hand side of \ceq{eq:F-asympt}. 

\begin{figure}
\begin{center}
\includegraphics[width=0.5\textwidth,trim={0 1.2cm 0 1.5cm},clip]{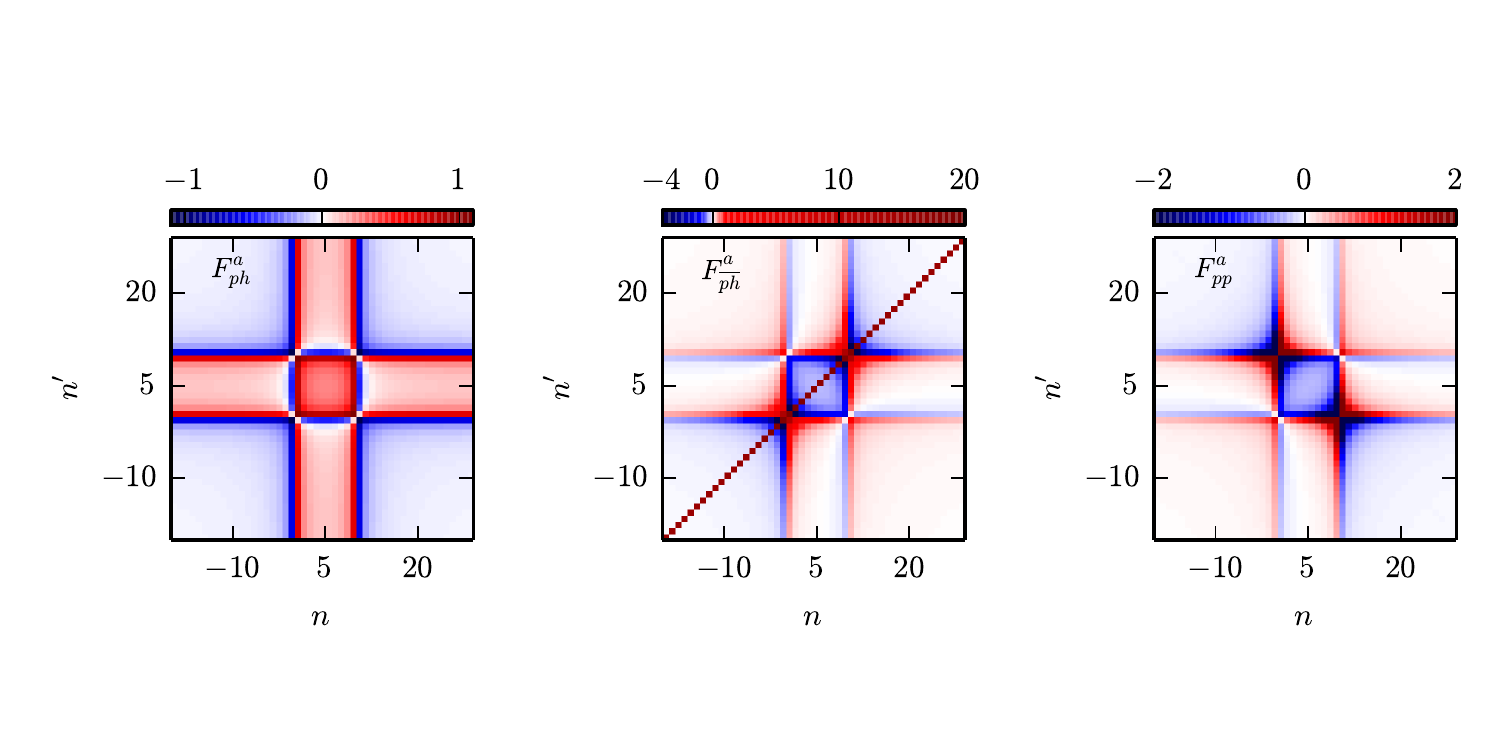}
\end{center}
\caption{\label{fig:vertex-assembly-atomic}The $ph$-, $\overbar{ph}$- and $pp$-parts of the asymptotic vertex $F_\ud^{\mathrm{asympt},\nu\nu'\omega_{10}}-U$ in $ph$-notation
  at $U=2$ and $\beta=8$.}
\end{figure}

Having at our disposal the asymptotic vertex, it is now possible to calculate how it deviates
from the complete vertex, similarly as it was done with the numerical data of the Hubbard model
and SrVO$_3$ above. 
Since the explicit analytical form of the asymptotic vertex is rather lengthy,
we print only the difference $R=F-F^{\mathrm{asympt}}$, which is, however, of much greater interest:
\begin{widetext}
\begin{equation}
\label{eq:R_ud}
R_\ud^{\nu_1 \nu_2 \nu_3 \nu_4} = \frac{1}{\nu_1 \nu_2 \nu_3 \nu_4} \left[ -\frac{3U^5}{16} 
+ \frac{\beta U^6}{64}\left(f\left(-\frac{U}{2}\right)-f\left(\frac{U}{2}\right)\right)\delta_{\nu_1\nu_2}\\
+ \frac{\beta U^6}{32}\left(-\frac{U}{2}\right)\delta_{\nu_1\nu_4}
- \frac{\beta U^6}{32}2f\left(\frac{U}{2}\right)\delta_{-\nu_1\nu_3} \right].
\end{equation}
\end{widetext}
Furthermore, we have
\begin{equation}
R_\uu^{\nu_1 \nu_2 \nu_3 \nu_4} = \frac{\beta U^6}{64} \frac{\delta_{\nu_1 \nu_4}-\delta_{\nu_1,-\nu_3}}{\nu_1 \nu_2 \nu_3 \nu_4}
\end{equation}
and
\begin{equation}
R_\udb^{\nu_1 \nu_2 \nu_3 \nu_4} = -R_\ud^{\nu_1 \nu_4 \nu_3 \nu_2}
\end{equation}
for the other spin-components. Slices of the purely asymptotic vertex $F^{\mathrm{asympt}}_\ud$ and the difference to the full vertex $R_\ud$ 
are shown in \cfg{fig:r-atomic}. We observe that indeed the differences of the full and asymptotic vertices
go to zero with $1/(\nu_1 \nu_2 \nu_3 \nu_4)$ for all components, meeting our initial requirement. 
Together with the delta-functions, \ceq{eq:R_ud} also motivates the asymptotic replacement condition
\ceq{eq:replacement-condition}.

\begin{figure}
\includegraphics[width=0.5\textwidth]{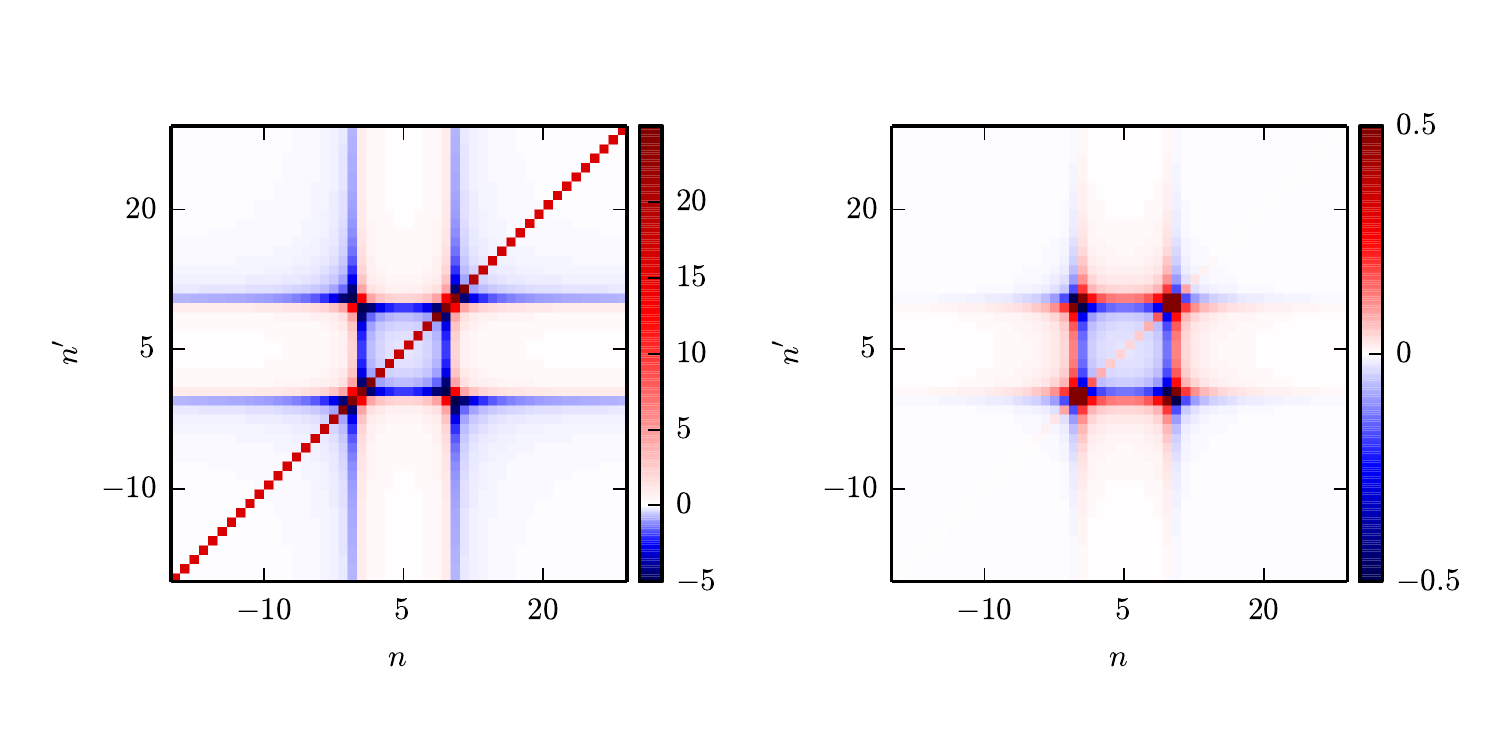}
\caption{\label{fig:r-atomic}$F_\ud^{\mathrm{asympt,\nu\nu'\omega_{10}}}$ (left) and $R_\ud^{\nu\nu'\omega_{10}}$ (right) in $ph$-notation at $U=2$ and $\beta=8$.}
\end{figure}

\begin{table}
\begin{center}
\bgroup
\def\arraystretch{2}
\begin{tabular}{l | c | c}
$K^{(1),\ell,\omega}$ & $pp$ & $ph$ \\ \hline
$\uparrow\uparrow$ & 
  $0$ & 
  $-\frac{\beta U^2}{4}\delta_{\omega 0}$ \\
$\uparrow\downarrow$ & 
  $-\frac{\beta U^2}{2} f\left(\frac{U}{2}\right) \delta_{\omega 0}$ & 
  $\frac{\beta U^2}{4} \left[ f\left(-\frac{U}{2}\right) - f\left(\frac{U}{2}\right)\right] \delta_{\omega 0} $ \\
$\overbar{\uparrow\downarrow}$ & 
  $\frac{\beta U^2}{2} f\left(\frac{U}{2}\right) \delta_{\omega 0}$ & 
  $-\frac{\beta U^2}{2} f\left(-\frac{U}{2}\right) \delta_{\omega 0}$
\end{tabular}
\egroup
\end{center}
\caption{\label{tab:k1}Kernel functions $K^{(1)}$ in particle-particle and particle-hole channel.
Frequencies are given in the channel-specific notations, see Appendix \ref{sec:freq-not}.}
\end{table}

\begin{table}
\begin{center}
\bgroup
\def\arraystretch{2}
\begin{tabular}{l | c | c}
$K^{(2),\ell,\nu\omega}$ & $pp$ & $ph$ \\ \hline
$\uparrow\uparrow$ & 
  $0$ & 
  $K^{(1),ph,\omega}_\uu \frac{U^2}{4\nu^2}$ \\
$\uparrow\downarrow$ & 
  $\frac{U^2}{4\nu(\nu-\omega)}(K^{(1),pp,\omega}_\ud-U)$ & 
  $\frac{U^2}{4\nu(\nu-\omega)}(K^{(1),ph,\omega}_\ud-U)$ \\
$\overbar{\uparrow\downarrow}$ & 
  $-\frac{U^2}{4\nu(\nu-\omega)}(K^{(1),pp,\omega}_\ud-U)$ & 
  $\frac{U^2}{4\nu(\nu-\omega)}(K^{(1),ph,\omega}_\udb+U)$
\end{tabular}
\egroup
\end{center}
\caption{\label{tab:k2}Kernel functions $K^{(2)}$ in particle-particle and particle-hole channel.
Frequencies are given in the channel-specific notations, see Appendix \ref{sec:freq-not}. }
\end{table}

\FloatBarrier
\bibliography{bibliography} 
\vfill\eject

\end{document}